\documentclass[twocolumn,showpacs,preprintnumbers,amsmath,amssymb,aps,prb]{revtex4-1}

\usepackage{graphicx}
\usepackage{dcolumn}
\usepackage{bm}

\begin{document}

\title{
Orbital entanglement and electron localization in quantum wires}
\author{Alberto Aleta}
\author{H\'ector Villarrubia-Rojo}
\affiliation{Departamento de F\'isica Te\'orica, Universidad de Zaragoza, Pedro Cerbuna 12, E-50009 Zaragoza, Spain }
\author{Diego Frustaglia}
\affiliation{Departamento de F\'isica Aplicada II, Universidad de Sevilla, E-41012 Sevilla, Spain}
\author{V\'ictor A. Gopar}
\affiliation{Departamento de F\'isica Te\'orica and BIFI, Universidad de Zaragoza, Pedro Cerbuna 12, E-50009 Zaragoza, Spain}

\date{\today}

\begin{abstract}

We study the signatures of disorder in the production of orbital electron entanglement in quantum wires.
Disordered entanglers suffer the effects of localization of the electron wave function and random fluctuations in entanglement production. This manifests in the statistics of the concurrence, a measure of the produced two-qubit entanglement. We calculate the concurrence 
distribution as a function of the disorder strength within a random-matrix approach.  We also identify significant constraints on the entanglement production as a consequence of the breaking/preservation of time-reversal symmetry.  Additionally, our theoretical results are independently supported by  simulations of disordered quantum wires based on a tight-binding model. 
 
\end{abstract}

\pacs{03.67.Bg, 73.23.-b, 72.15.Rn, 73.63.Nm}

\maketitle

\section{Introduction}

Quantum entanglement (nonclassical correlations between separated partners) is identified as a key resource for emerging information technologies \cite{NC-book,review_andreas}, especially in modern quantum electronics. In addition, the concept has been remarkably useful to shed new light on well-studied fields such as mesoscopic transport. 
Electron entanglement can be produced either by interacting mechanisms\cite{recher,lesovik,oliver,samuelsson_inter, bouchiat,saraga} (e.g., exchange coupling and superconducting pairing) or noninteracting ones \cite{bose,beenakkker-emary,samuelsson,signal,lebedev,FC09,liliana} (based on exchange correlations in scattering processes from external potentials). Electronic devices such as quantum dots and quantum wires have been proposed to produce entanglement of electrons without interactions \cite{B06,B07}. 
The efficiency of these noninteracting entanglers depends on the  scattering of electrons  traveling  through the system. Thus, features of the scattering 
matrix $S$ associated with the entangler, such as symmetries and dimensionality, determines the degree of electron entanglement.

Effects of quantum chaotic scattering on orbital entanglement production in quantum dots have been studied in the past \cite{BKMY04,B06, FMF06, gopar_frustaglia, rodriguez,almeida,vivo}. In these ballistic microstructures, electrons are elastically scattered, while the chaotic character of the scattering produces stochastic fluctuations of the orbital entanglement. Therefore a statistical analysis of the  entanglement is required. 
 In essence, those works addressed the effect of the underlying  classically chaotic dynamics of the dot on entanglement production.\cite{bambi}
 
 In contrast to the ballistic scattering in quantum dots, in which electrons are scattered off the dot boundary, in disordered quantum wires  electrons suffer multiple scattering, e.g., from impurities. Thus, if a quantum wire is used as entangler, a new ingredient is expected to play a relevant role in the description of the properties of the electron  entanglement: the localization length of the electron wave functions,  determined by the disorder strength. 
 Actually, very recent experiments \cite{DJPRCJWRG13} have opened the possibility of producing entangled electron  pairs in quantum conductors  from single excitations, named levitons, in the Fermi sea \cite{levitov_1,levitov_2}.

In general, disorder effects in quantum electronic devices have been of interest  from both fundamental and applied points of view. For instance, the presence of disorder in a system leads to the widely studied phenomenon of Anderson localization, a spatial localization of the electron wave function.  Lattice defects and impurities are two examples of sources of disorder,  which may be  unavoidable in electronic systems. Therefore, 
disorder effects in phase-coherent quantum transport have been of particular interest. In addition, the presence of disorder 
gives a stochastic character to the electron scattering processes and  calls for a statistical analysis of scattering-dependent phenomenon, such 
as the production of orbital entanglement.

Random-matrix theory has been successfully applied to study different statistical properties of  scattering in disordered systems \cite{carlo-review,mello_book}. In particular, by using a  scattering approach to the problem of quantum  transport (Landauer-B\"uttiker approach)  several  electronic properties of disordered quantum wires have been analyzed within a random-matrix approach.  In this theoretical framework, a Fokker-Planck equation for the probability density of the transmission eigenvalues of a quantum wire of length $L$ and 
width $W$  has been derived \cite{mello_book}. It turns out that this transmission probability   depends on only a single  physical parameter: the localization length. 

In this work, the statistical properties of the production of orbital entanglement in disordered quantum wires are studied. We adopt a random-matrix approach to study the distribution of an entanglement indicator: the two-qubit \emph{concurrence}, $\mathcal{C}$. We show the evolution of the concurrence distribution as a function of the disorder strength for different symmetry classes: broken and preserved time-reversal symmetry (TRS). As we show below, the presence of TRS is crucial for the production of highly entangled states.  In general, as the strength of disorder decreases, the possibility of having highly entangled states increases. However, for entanglers with broken TRS the probability distribution of concurrence vanishes at maximum entanglement ($\mathcal{C}=1$), for  any value of the strength of the disorder. All our analytical results are supported by independent numerical simulations based on a standard tight-binding Hamiltonian model.
\begin{figure}
\begin{center}
\includegraphics[width=0.9\columnwidth]{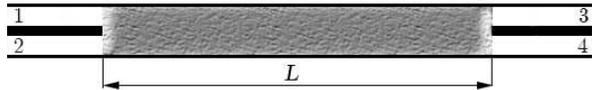}
\caption{Setup of a quantum wire entangler of length $L$. 
The left (1 and 2) and right (3 and 4) ideal leads are attached to a disordered region [shaded (gray)]. An electron leaving the quantum wire  to the left (right) side 
can escape through the perfect lead 1 or 2 (3 or 4), defining a two-level quantum system.}
\label{fig_1}
\end{center}
\end{figure}

This paper is organized as follows: In Sec. \ref{setup} we introduce the setup, followed by a brief discussion of noninteracting entanglement production and a precise definition of the concurrence. 
In Sec. \ref{model}, we elaborate on the statistics of transmission eigenvalues in disordered systems to calculate the concurrence distribution.  After a brief introduction of our numerical model, in Sec. \ref{comparison} we present the results of our simulations and compare them with the theoretical predictions obtained in the previous section. Finally, a closing summary and discussion are given in Sec. \ref{conclusions}.

\section{Entangler setup \label{setup}}

Our setup is sketched in Fig.~\ref{fig_1}. Two single-channel leads are attached at the ends of the wire. Left and right  leads are  
connected to  electron reservoirs $\mu_L$ and $\mu_R$, respectively. 
This resembles the orbital entanglers proposed in Refs. \cite{BKMY04,FMF06,gopar_frustaglia}, provided the chaotic quantum dot is replaced by a disordered quantum wire. A low bias voltage between reservoirs leads to a 
coherent current along the wire from left to right. Exchange correlations due to electron scattering within the 
wire create the conditions for the production of entanglement between transmitted (to the right) and reflected (to the left) electrons, as we show in the following. 

We start by considering an uncorrelated two-particle state incoming from the left reservoir 
in Fig.~\ref{fig_1}:
\begin{equation}
|\Psi_{\rm in}\rangle=a_1^\dag a_2^\dag |0\rangle.
\label{Psi-in}
\end{equation}
The $a_i^\dag$ creates an incoming electron excitation in lead $i=1, 2$ above the Fermi sea $|0\rangle$ at 
zero temperature. For simplicity, we disregard spin degeneracy (equivalently, one can consider spin-polarized incoming electrons). Let $S$ be the wire's scattering matrix relating incoming and 
outgoing states. In general, the $S$-matrix can be written as
\begin{equation}
S=\left(
\begin{array}{cc}
r &  t'   \\
t & r'
\end{array}
\right)\; ,
\label{S}
\end{equation}
where $r,r',t,$ and $t'$ are $2 \times 2$ reflection and transmission matrices, respectively. In the presence of 
time reversal symmetry (TRS), $S$ is unitary and symmetric. Instead, for broken TRS (due to, e.g., the application of a magnetic flux) the $S$-matrix is only unitary.

The outgoing state $|\Psi_{\rm out}\rangle$ is a coherent superposition 
of orbital channels determined by the single-particle scattering matrix $S$. $|\Psi_{\rm out}\rangle$   can be split into three components \cite{WV03} representing sectors of the Fock space with 
different local particle numbers at the left ($n_{\rm L}$) and right ($n_{\rm R}$) ends of the wire 
such that the total particle number remains constant ($n_{\rm L}+n_{\rm R}= 2$):
\begin{equation}
|\Psi_{\rm out}\rangle=\sum_{n_{\rm L},n_{\rm R}} |n_{\rm L},n_{\rm R}\rangle=|2,0\rangle+|0,2\rangle+|1,1\rangle. 
\label{Psi-out-n}
\end{equation}  
The only sector contributing to the orbital entanglement is the one with equal occupancy at both ends of the wire. This is given by $|1,1\rangle = \sum_{pq} (r_{p1} t_{q2}-t_{q1} r_{p2}) b_p^\dag b_q^\dag |0\rangle$, 
with $p = 1,2$ and $q=3,4$, where $b_j^\dag$ creates an outgoing electron excitation in lead $j=1,...,4$.
The $|2,0\rangle$ and $|0,2\rangle$, instead, are separable in terms of the bipartition left-right \cite{BKMY04,FMF06,B06,gopar_frustaglia}.
An electron leaving the quantum wire to the left side can choose between lead $1$ and lead $2$ for escaping (see Fig.~\ref{fig_1}). 
This defines a two-level quantum system or qubit. The same happens with an electron escaping to the right side through leads $3$ and $4$. 
As a consequence, we conclude that the component $|1,1\rangle$ in Eq. (\ref{Psi-out-n}) corresponds (up to a normalization factor) to a two-qubit entangled state. 

An efficient measure for quantification of two-qubit entanglement is the concurrence $\mathcal C$. This is defined as\cite{W98}
\begin{equation}
\mathcal C(\rho)\equiv {\rm max}\{0,\lambda_1-\lambda_2-\lambda_3-\lambda_4\}.
\label{C}
\end{equation}
The $\lambda_i$'s are the eigenvalues (in decreasing order) of the matrix $\rho \tilde{\rho}$, where $\rho$ is a $4\times 4$ two-qubit density matrix ($\rho=|1,1\rangle \langle1,1|/\langle1,1|1,1\rangle$ in our case) and  
$\tilde{\rho}=(\sigma_y \otimes \sigma_y)\rho^*(\sigma_y \otimes \sigma_y)$, with $\sigma_y$  the second Pauli matrix. The concurrence runs from 0 to 1, corresponding to separable and maximally entangled (Bell) states, respectively. Those states with $0 < \mathcal C < 1$ are partly entangled states. It turns out that 
the concurrence is determined by the scattering amplitudes, and can be written in terms of the transmission eigenvalues $\tau_1$ and $\tau_2$ 
of the  product $tt^\dagger$  as\cite{BKMY04,B06}
\begin{equation}
\label{concu_t1t2}
\mathcal C=\frac{2\sqrt{\tau_1(1-\tau_1)\tau_2(1-\tau_2)}}{\tau_1+\tau_2-2\tau_1\tau_2} .
\end{equation}
Note that the entanglement maximizes ($\mathcal C=1$) for $\tau_1 =\tau_2$, 
and minimizes ($\mathcal C=0$) for $\tau_1=0$ and $\tau_2=1$ or $\tau_1=1$ and $\tau_2=0$. 

As we have mentioned, the presence of disorder in the entangler gives a stochastic character to the scattering processes and therefore to the transmission. Thus, from Eq. (\ref{concu_t1t2}), the statistics of the concurrence is determined by the statistical properties of the transmission eigenvalues.

\section{Statistics of transmission eigenvalues and concurrence \label{model}}

Several statistical properties of the transmission eigenvalues of the transfer matrix $tt^\dagger$  have been investigated in disordered quantum wires within a scaling theory of localization. By considering a disorder wire of length $L$ and width $W$ with perfect leads attached at the ends (each one supporting $N$ transverse modes or channels), it has been found that the 
distribution probability of the transmission eigenvalues $P(\tau_1, \tau_2,\ldots, \tau_N)$ is determined by
a Fokker-Planck equation. This diffusion equation, also known as the Dorokhov-Mello-Pereyra-Kumar (DMPK) equation, is an evolution equation for $P(\tau_1, \tau_2,\ldots, \tau_N)$ as a function of the length of the system \cite{dmpk, mello_book}.  Exact solutions of this equation are known for Hamiltonian systems with broken TRS\cite{rejaei,rejaei_prb} ($\beta=2$). When the invariance is preserved ($\beta=1$), instead,  only approximated solutions are known for the insulating and metallic regimes.

To calculate the concurrence distribution we use an expression for $P_\beta(\tau_1,\tau_2, \ldots, \tau_N)$ which has been shown to be a good approximation from the metallic to the insulating regimes for both $\beta = 1$ and 2 symmetries.\cite{crossover,annals,gopar_Pb} Actually, to study the concurrence, Eq. (\ref{concu_t1t2}), we need the joint distribution of the transmission eigenvalues $\tau_1$ and $\tau_2$, only.
For convenience, we introduce a change of variables, $\tau_i=1/\cosh^2 x_i$. Thus,  the joint distribution  $P_\beta(x_1,x_2)$ reads 
\begin{eqnarray}
&& P_\beta(x_1,x_2)=\mathcal{N}(s)|(\sinh^2 x_1-\sinh^2 x_2)(x_1^2-x_2^2)|^{\beta/2}   \nonumber \\
 & &\times \prod_{i=1}^2 \left[\exp(-(\beta+2) x_i^2/2s)(x_i\sinh 2x_i)^{1/2}\right],
\end{eqnarray}
where $s=L/l$ is the length of the system in units of the mean free path $l$ and $\mathcal{N}(s)$ is a normalization constant. We note  that 
the complete statistics of the transmission eigenvalues is determined by the sole  parameter $s$.

We are now ready to calculate the concurrence distribution $P_\beta(\mathcal C)$. By implementing the change of variables $\tau_i \to x_i$ in  Eq. (\ref{concu_t1t2}), the concurrence distribution is determined by
\begin{eqnarray}
\label{pofC}
P_\beta (\mathcal C)&=& \int \int \text{d}x_1  \text{d}x_2   P_\beta(x_1,x_2) \nonumber \\
&&\times \delta\left[ \mathcal C-\frac{2\sinh x_1\sinh x_2}
{\sinh^2x_1 +\sinh^2 x_2} \right] .
\end{eqnarray}
One of the integrals in Eq.~(\ref{pofC}) can be performed analytically, finding 
\begin{equation}
\label{pofCx2}
P_\beta(\mathcal C)=\mathcal{N}(s) \int_0^\infty \left[ g\left( r_+(x_2)\right)+g\left(r_-(x_2)\right) \right] \text{d}x_2  ,
\end{equation}
where we have defined $r_{+}$ and  $r_{-}$ as  
\begin{equation}
\label{xpm}
 r_{\pm}(x_2) = \sinh^{-1}(K \sinh x_2) 
\end{equation}
with 
\begin{equation}
\label{K}
 K =\frac{1\pm \sqrt{1-\mathcal C^2}}{{\mathcal C}}  .
\end{equation}
Also, the function $g(r)$ in Eq.~(\ref{pofCx2}) is defined as 
\begin{eqnarray}
\label{g}
g(r_{\pm}&)=&\sqrt{K} \left( K^2+1 \right)^2 \left| \left( K^2-1 \right) \right|^{(\beta-2)/2} \nonumber \\
&\times &\sqrt{r_\pm x_2}\left| x_2^2-r_\pm^2 \right|^{\beta/2}\exp \left[-\frac{\beta +2}{2s}\left( r_\pm^2+x_2^2 \right)\right] \nonumber \\
&\times &  \left( \frac{\cosh^2 x_2}{1+K^2 \sinh^2 x_2} \right)^{1/4} \sinh^{2+\beta} x_2  .
\end{eqnarray}
Thus, the concurrence distribution can be obtained by numerical integration of Eq.~(\ref{pofCx2}) for any given value of $s$. In the ballistic limit ($s \ll 1$), however,  we can analytically perform the integrals in Eq. (\ref{pofC}), finding  simple expressions for both symmetry classes $\beta=1$ and 2:  
\begin{equation}
\label{pofc1}
P_1 (\mathcal C) = 2\mathcal C
\end{equation}
for entanglers with preserved TRS,  and 
\begin{equation}
\label{pofc2}
P_2 (\mathcal C) = 3\mathcal C \sqrt{1-\mathcal C^2} 
\end{equation}
for entanglers with broken TRS. 
From the above distributions, Eqs. (\ref{pofc1}) and (\ref{pofc2}),  we readily obtain the first moments (mean values and variances) of the concurrence in the ballistic limit: 
\begin{equation}
\label{averageC}
\langle \mathcal C \rangle =\left\{
\begin{array}{ll}
2/3   &  \mbox{for $\beta=1$,}\\
3\pi/16  & \mbox{for $\beta=2$}, \\
\end{array}
\right.
\end{equation}
while for the variances we have
\begin{equation}
\label{varianceC}
\mathrm{var}(\mathcal C) =\left\{
\begin{array}{ll}
 1/18    &  \mbox{for $\beta=1$,} \\
  2/5-9\pi^2/256   & \mbox{for $\beta=2$ .} 
\end{array}
\right.
\end{equation}
Interestingly, we point out  that the average concurrence, Eq. (\ref{averageC}), in the ballistic  limit is much higher than in quantum 
dot entanglers,\cite{average} in which electrons undergo  ballistic chaotic scattering.

In the next section, we discuss and verify our results [Eqs. (\ref{pofCx2}), (\ref{pofc1}), and (\ref{pofc2})] with the support of independent numerical simulations based on a tight-binding model.


\section{Numerical simulations }
\label{comparison}

We now introduce a  numerical model to verify the theoretical predictions in the previous section.  
We consider the standard  tight-binding Hamiltonian  given by 
\begin{equation}
 H=\sum_i \epsilon_i c_i^\dagger c_i -  \sum_{<ij>}(t_{ij}c_i^\dagger c_j +\text{h.c.}),
\end{equation}
where $\epsilon_i$ is the onsite energy,  $t_{i,j}$ represents the hopping element between nearest-neighbor sites, and $c_i^\dagger$ ($c_i$) is the creation (annihilation) operator for electron excitations at site $i$. In this model, the disorder is implemented by random onsite  energies, sampled  from a constant 
distribution in the interval $[-w/2,w/2]$. In this paper, the statistics of the concurrence are collected from 20000  different disorder realizations. 

Disordered quantum wires can be  characterized by  the average dimensionless conductance $\langle G \rangle$, a standard quantity in quantum transport. We recall that, within the scattering approach to electronic transport,  the  dimensionless conductance $G$ is given in terms of the transmission matrix $t$ by $G=\text{trace}(tt^\dagger)$.  In our numerical simulations, we can produce an ensemble of quantum wires with a desired value of $\langle G \rangle$ by controlling the degree of disorder $w$ ($\langle G \rangle$ is a decreasing function of the disorder strength). Thus, we 
compare the concurrence distributions from the numerical simulations with the corresponding theoretical predictions, both distributions having the same value for the  average conductance.

\subsection{Wire entangler with TRS}

\begin{figure}
\includegraphics[width=\columnwidth]{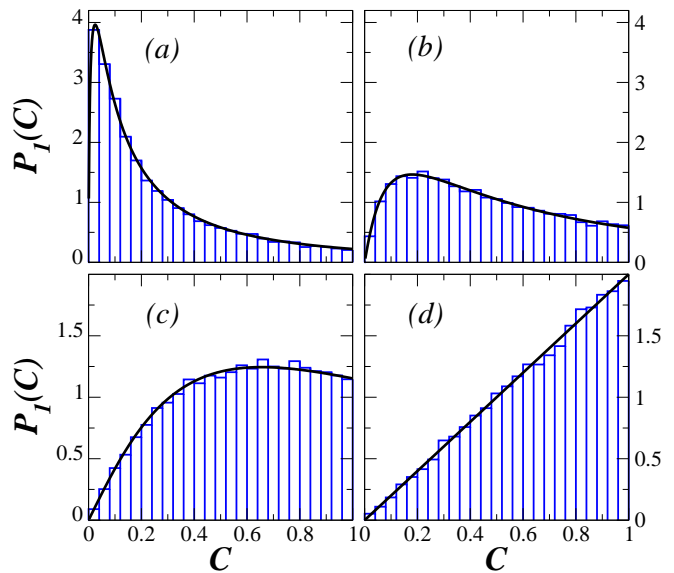}
\caption{(Color online) Concurrence distributions for quantum wires with time-reversal symmetry ($\beta=1$). The strength of the disorder 
decreases from (a) to (d): $w=2.01, 1.36, 0.86,$ and $ 0.20$, respectively. From (a) to (d), the values of the average conductance are $\langle G \rangle=0.2, 0.5, 1.0,$ and 1.9. 
Solid lines in (a), (b), and (c) were obtained from Eq. (\ref{pofCx2}) with $s = 5.4, 2.4$, and 0.9, respectively, while the solid line in (d) 
was computed from Eq. (\ref{pofc1}). Histograms 
were obtained from numerical simulations. A good agreement between theory and 
numerics is seen in all cases.
}
\label{fig_2}
\end{figure}

First, we consider the case of quantum wires with preserved TRS, symmetry class $\beta = 1$.
In Fig. \ref{fig_2}, we show the concurrence distribution for different values of  the disorder strength $w$. 
The histograms correspond to the distribution obtained from the numerical simulations and the solid lines are the theoretical predictions.
In Figs. \ref{fig_2}(a) to \ref{fig_2}(d), the degree of disorder is such that  $\langle G \rangle= 0.2,0.5,1.0,$ and 1.9, respectively. 
We see in Fig. 2 that large values of the concurrence ($\mathcal{C}\approx 1$) are statistically favored as the disorder decreases [i.e., from Fig. \ref{fig_2}(a) to Fig. \ref{fig_2}(d)]. In particular, in Fig. \ref{fig_2}(d), we have plotted  our  analytical expression, Eq. (\ref{pofc1}), valid in the ballistic regime limit. In all cases, the agreement between numerics and theory is very good. We, further, note the possibility 
that maximally entangled states are produced in all cases [note that $P_1(\mathcal{C})$ is finite at $\mathcal{C}=1$]. 

\subsection{Wire entangler with broken TRS}

We implement the symmetry class $\beta=2$ (broken TRS) by introducing  in our tight-binding calculations 
a magnetic field perpendicular to the wire. 

\begin{figure}
\includegraphics[width=\columnwidth]{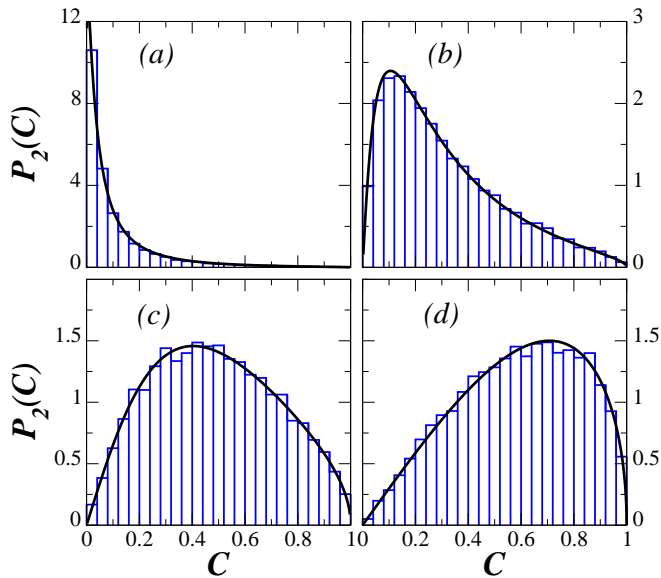}
\caption{(Color online) Concurrence distributions for quantum wires with broken time-reversal symmetry ($\beta=2$). The strength of the 
disorder 
decreases from (a) to (d): $w=2.05, 1.31, 0.76,$ and $0.15$, respectively. From (a) to (d), the values of the average conductance are $\langle G \rangle=0.2, 0.5, 1.0,$ and 1.9. 
Solid lines in (a), (b), and (c) were obtained from Eq. (\ref{pofCx2}) with $s=6.5, 2.7$ and 1.0, respectively, while the solid line in (d) was computed from Eq. (\ref{pofc2}). Histograms were obtained from the tight-binding numerical simulations with a perpendicular magnetic field: (a), (b), (c) $eBa^2/\hbar=1$ ($a$ being  the lattice spacing), and (d) $eBa^2/\hbar=0.5$. For all cases we have  a good agreement between theoretical  and numerical results.}
\label{fig_3}
\end{figure}

In Fig. \ref{fig_3} we show the  concurrence distributions for different degrees of disorder, organized as in Fig. \ref{fig_2}, from strong to weak strength of disorder. In order to see the effects of breaking the TRS, 
the values of the strength disorder in Fig. \ref{fig_3} are such that the values of the average conductance are the same as in the previous case ($\beta=1$), i.e., $\langle G \rangle= 0.2,0.5, 1.0$ and 1.9.  
As we can see, once again, a very good agreement is found between numerical results (histograms) and theoretical predictions (solid lines) in all cases. 

Similarly to the case $\beta=1$, in Fig. \ref{fig_3} we find that the statistical distribution favors largely entangled states for weakly disorder wires. In spite of this, the production of maximally entangled states appears now to be forbidden due to the broken TRS [$P_2(\mathcal{C})$ vanishes at $\mathcal{C}=1$ in all panels in Fig. \ref{fig_3}], in contrast to the symmetry class $\beta=1$, in which $P_1(\mathcal{C})$ is finite at $\mathcal{C}=1$.  
This difference between $\beta = 1$  and $\beta= 2$ symmetries resembles those observed in chaotic dot entanglers \cite{gopar_frustaglia}.

\section{summary and conclusions \label{conclusions}}

Effects of disorder in quantum electronic devices plays a central role. For instance, the presence of disorder 
in a quantum wire produces the spatial localization of the electron wave function--Anderson localization--and gives  a random character to 
the electron scattering. As a consequence, a statistical analysis of physical quantities that depend on scattering  processes is required.

Here, we have addressed the production of two-qubit orbital entanglement in disordered quantum wires. The concurrence is a measure of the degree of entanglement,  which depends on the scattering matrix through the  transmission eigenvalues. Since the disorder strength determines the degree of localization of the electron wave functions, it is  expected that the entanglement production is affected too.  
We have used a random-matrix approach to study the statistical properties of the concurrence. 
This theoretical framework has been used to investigate some statistical properties of phase-coherent transport,  
such as the conductance through quantum wires.  The concurrence is, however,  a more complex quantity than the conductance, in the sense that it involves correlations between the transmission eigenvalues $\tau_1$ and $\tau_2$, which are not present in the conductance. We thus have  further studied the validity of the random matrix approach to quantum phenomena. 

We have analytically calculated  the complete concurrence distribution of a quantum-wire entangler,  for both symmetries: broken and 
preserved  TRS. Effects of different degrees of electron localization are revealed by showing the evolution of 
the concurrence distribution with the strength of the disorder. We have  
found that disorder statistically hinders the production of highly entangled states. However, the possibility of producing maximally entangled states is fully determined 
by TRS, independently of disorder. This coincides with previous results on orbital entanglement production in chaotic quantum dots, \cite{gopar_frustaglia},  although the concurrence distribution of chaotic and disordered  entanglers are completely different. Thus, TRS appears as a fundamental ingredient to produce maximal orbital entanglement.  Also, by comparing the concurrence averages  of quantum wires and quantum dots, it is interesting to  notice that 
quantum wires  in the ballistic regime  [system length much shorter than the mean free path;  Figs. \ref{fig_2}(d) and \ref{fig_3}(d)] can produce, on average, larger orbitally entangled states than in quantum dots with (ballistic) chaotic scattering. Additionally, our theoretical results have been  supported by numerical simulations implemented in a tight-binding model, showing an excellent agreement. 
 
Finally, we note that our predictions could be evaluated by implementing multiple sources of single-particle excitations (levitons). \cite{levitov_1,levitov_2} Using quantum wires, Dubois \emph{et al}. \cite{DJPRCJWRG13} have opened  the possibility of producing 
entangled  pairs by applying voltage pulses.  The degree of entanglement could thus be determined by 
introducing some kind of state tomography.\cite{witness_1,witness_2}

\acknowledgments

D. F. and V. A. G acknowledge support from MINECO (Spain) under Projects No. FIS2011-29400 and No. FIS2012-35719-C02-02, respectively.


\begin{thebibliography}{50}
\bibitem{average} For quantum-chaotic dots, it was found\cite{gopar_frustaglia} that $\langle \mathcal{C}\rangle_{\beta=1}=\ln 4-1\approx 0.386$ and $\langle \mathcal{C}\rangle_{\beta=2}=4\pi(21\pi/64-1)\approx 0.387$
\end{thebibliography}
\end{document}